\documentclass{osa-article}

\journal{boe}
\usepackage{placeins}
\usepackage{hyperref}

\articletype{Research Article}

\begin{document}

\title{Imaging human blood cells in vivo with oblique back-illumination capillaroscopy}

\author{Gregory N. McKay,\authormark{1} Nela Mohan,\authormark{1} and Nicholas J. Durr\authormark{1,*}}

\address{\authormark{1}Department of Biomedical Engineering, Johns Hopkins University (JHU), 3400 N. Charles Street, Baltimore, MD, 21218, USA\\}

\email{\authormark{*}ndurr@jhu.edu} 
\homepage{http://durr.jhu.edu} 



\begin{abstract}
We present a non-invasive, label-free method of imaging blood cells flowing through human capillaries \textit{in vivo} using oblique back-illumination capillaroscopy (OBC). Green light illumination allows simultaneous phase and absorption contrast, enhancing the ability to distinguish red and white blood cells. Single-sided illumination through the objective lens enables 200 Hz imaging with close illumination-detection separation and a simplified setup. Phase contrast is optimized when the illumination axis is offset from the detection axis by approximately 225 $\mu$m when imaging $\sim$80 $\mu$m deep in phantoms and human ventral tongue. We demonstrate high-speed imaging of individual red blood cells, white blood cells with sub-cellular detail, and platelets flowing through capillaries and vessels in human tongue. A custom pneumatic cap placed over the objective lens stabilizes the field of view, enabling longitudinal imaging of a single capillary for up to seven minutes. We present high-quality images of blood cells in individuals with Fitzpatrick skin phototypes II, IV, and VI, showing that the technique is robust to high peripheral melanin concentration. The signal quality, speed, simplicity, and robustness of this approach underscores its potential for non-invasive blood cell counting. 
\end{abstract}

\section{Introduction}

The complete blood count (CBC) is an invaluable diagnostic test that provides clinical insight in nearly every disease state. The CBC is the most frequently ordered blood test, with approximately 34.5 million Medicare-reimbursed CBCs performed annually in the United States \cite{Mph2014}. While generally effective, the CBC requires invasive blood draw followed by laboratory analysis. For some vulnerable patient populations, such as immunocompromised cancer patients at risk for hospital-acquired infection and neonates at risk of phlebotomy-induced anemia, patient management could be dramatically improved by a non-invasive CBC \cite{Tai2017,Crawford2003,Widness2008}.

Non-invasive imaging of human vasculature continues to reveal interesting, clinically relevant phenomena despite its long history \cite{Brown1925,Jung2013,Lefford1986,Chang1997,Curtis1999}. Recently, the frequency of optical absorption gaps between red blood cells imaged in nailfold capillaries has been inversely-correlated with neutropenia in cancer patients \cite{Bourquard2018}. However, due to the depth of these capillaries, individual white blood cells (WBCs) have not been resolved. Further, absorption of visible light by melanin in the epidermis results in a rapid degradation of image quality, making even optical absorption gaps difficult to capture when imaging the nailfold of individuals with Fitzpatrick skin phototype types V and VI. 

Several modern microscopy techniques have been investigated in recent years for imaging blood cells \textit{in vivo}. Methods such as reflectance confocal, second-, and third-harmonic generation microscopy have shown promise as research tools for \textit{in vivo} blood cell imaging, though all of these methods suffer from higher cost, increased complexity, and slower imaging speeds than capillaroscopy \cite{Tsai2002,Chen2012a,Rajadhyaksha1999}. Spectrally-encoded flow cytometry can distinguish between granulocyte and mononuclear cell populations in the human lip, though this technique requires a complicated spectral confocal microscope and the challenging alignment of intersecting an imaging line with a capillary in 3D space \cite{Golan2012a,Winer2017}. Recently, epi-illumination gradient light interference microscopy (epi-GLIM) demonstrated tomographic reconstructions of thick, scattering biological specimens, though at an imaging speed of only 4 Hz \cite{Kandel2019}. In 2012 Ford et al. introduced oblique back-illumination microscopy (OBM), a technique that enables phase contrast in thick, turbid media with epi-illumination, which was further developed for video-rate acquisition and quantitative reconstruction \cite{Ford2012,Ford2013,Obles2019}. The offset of illumination and detection axes produces an intensity gradient across the imaging field of view, and results in phase contrast due to the net oblique orientation of light backscattered through the focal plane. The ability to deploy absorption and phase contrast in an epi-illumination configuration has laid the foundation for the high-speed imaging of blood cells \textit{in vivo} with a simple optical setup, provided superficial capillaries are present. 

The choice of anatomical location is paramount for successful implementation of OBM for capillary imaging, with phase contrast extending only to a depth of approximately 100 $\mu$m in highly-scattering media like skin \cite{Ford2012}. Human capillaries in the nailfold, lying within dermal papillae just beyond the dermal-epidermal junction, are often 150-400 $\mu$m beneath the tissue surface \cite{Baran2015,Braverman1997}. In contrast, the epithelial lining of the oral mucosa is without a stratified squamous layer and can be less than 100 $\mu$m in thickness \cite{Presland2000,Prestin2012}. Further, melanocytes within the oral mucosa, and in particular the floor of the mouth and ventral tongue, tend to have fewer connections to adjacent keratinocytes, leading to lower melanin distribution as compared to peripheral skin \cite{Barrett1994, Feller2014}. Together, the superficiality of capillaries and decreased absorption due to reduced melanin distribution make the oral cavity an advantageous location for blood cell imaging in humans.

This paper demonstrates that phase contrast can be generated using oblique back-illumination in a simple capillaroscope by critically imaging an LED at a lateral offset with respect to the optical axis of the detector, rather than using an external fiber-based source \cite{Ford2012,Ford2013,Obles2019}. In addition to obviating the need for external fiber optics, this approach allows for small and tunable illumination-detection offset. Using green light and a 225 $\mu$m illumination-detection offset, simultaneous phase and hemoglobin absorption contrast is achieved, allowing clear imaging and differentiation of white and red blood cell types from a single image. With this approach, we demonstrate 200 Hz imaging of red blood cells, white blood cells, platelets, and sub-cellular granules when imaging ventral tongue capillaries of human participants with a wide range of Fitzpatrick skin types. The signal quality, speed, simplicity, and robustness of this approach may make it a practical technique for measuring complete blood counts non-invasively.

\section{Methods}

\subsection{Imaging System}
\begin{figure}[h!] 
\centering\includegraphics[width=12cm]{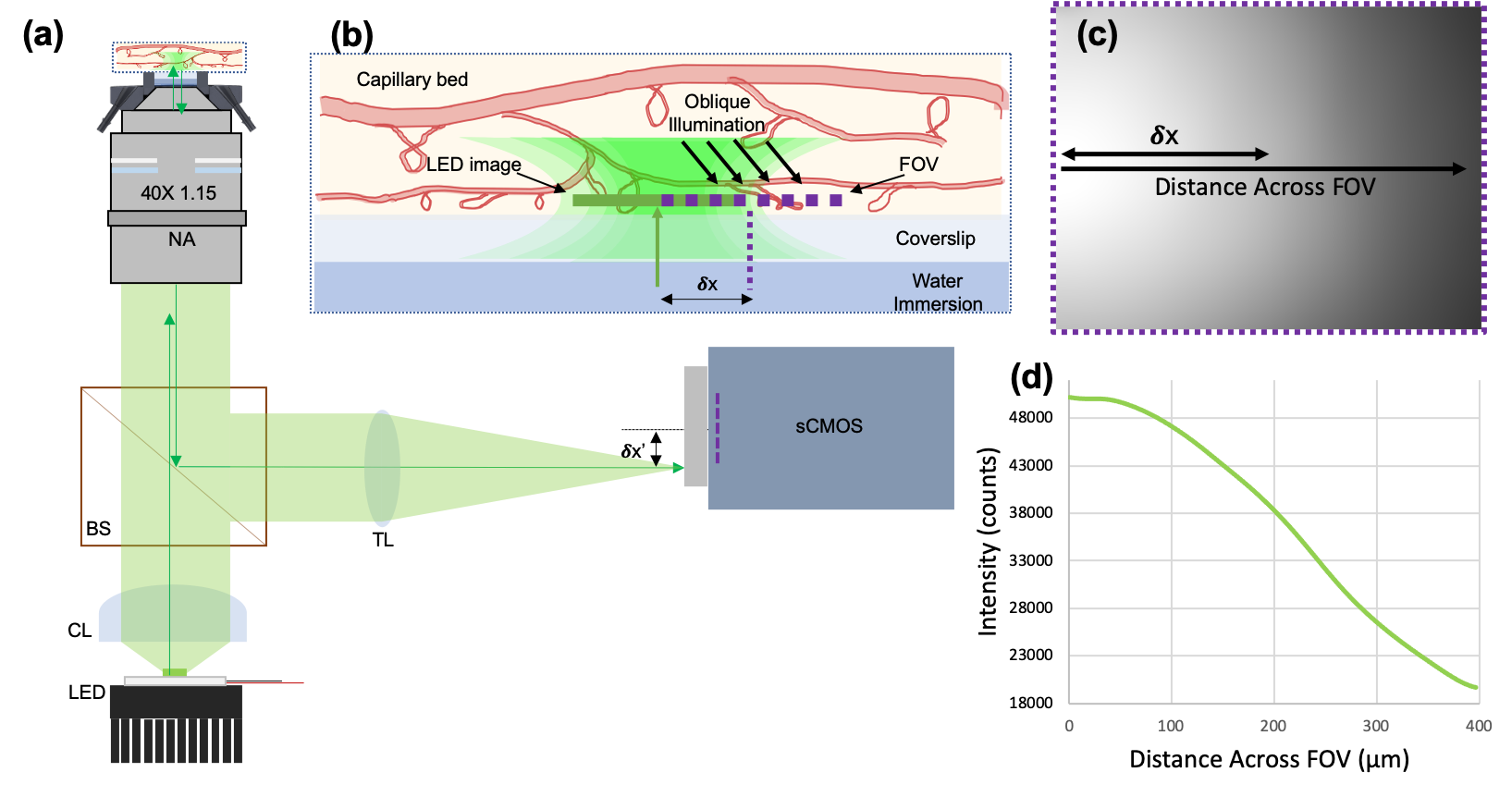}
\caption{(a) Oblique back-illumination capillaroscope optical layout. An LED is placed at the focal plane of a collimating lens (CL), and imaged critically into a capillary bed by a 40x objective lens. Scattered light is reflected off a beamsplitter (BS) and focused onto a laterally-displaced sCMOS by a tube lens (TL). (b) Oblique illumination is produced by offsetting the field of view (FOV) from the image of the LED. The camera displacement ($\delta x^{\prime}$) causes a 1/40x displacement in object space ($\delta x$). (c)-(d) An intensity gradient is produced, making the illumination-detection offset vary with distance across the full FOV for a fixed $\delta x$ = 208 $\mu$m.}
\label{fig:Figure1}
\end{figure}

Illumination is generated by a green LED (Superbright 1W XLamp LED, 88 lm, 527 nm) critically imaged into a capillary bed by a 20 mm collimating lens (Thorlabs ACL2520U-A) and 40x microscope objective (Nikon 1.15 NA APO LWD WI $\lambda$S). To produce phase contrast, the detector is laterally translated off-axis (Figure \ref{fig:Figure1}(a)-(b)). The offset between the illumination and detection axes results in a gradient of illumination intensity across the full field of view (FOV) (Figure \ref{fig:Figure1}(c)-(d)), and consequently, a net obliquity of light passing through blood cells \cite{Ford2012}. A 50:50 non-polarizing beamsplitter (CCM1-BS013) is used to redirect the light scattered by the tissue through a 200 mm tube lens (Thorlabs AC254-200-A-ML) and onto an image sensor (pco.edge sCMOS 5.5). 

\begin{figure}[h!] 
\centering\includegraphics[width=10cm]{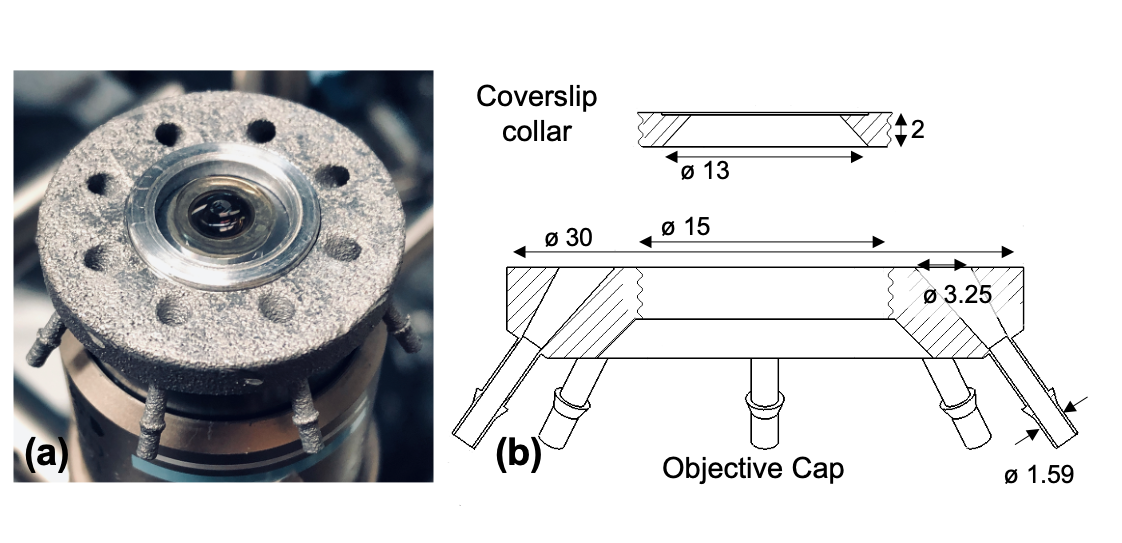}
\caption{(a) Image of pneumatic objective cap, and (b) mechanical drawing (mm units).}
\label{fig:Figure7}
\end{figure}

A custom, pneumatic objective cap was designed in Solidworks and 3D printed in aluminum (Figure \ref{fig:Figure7}). This cap contained eight suction ports, equally spaced every 45 degrees about a 21 mm diameter circle around a coverslip collar. The correction collar of the microscope objective was set to 150 $\mu$m for use with 12 mm diameter $\#$1 thickness coverslips. The imaging depth of the microscope was tuned to match the depth of the superficial capillaries by manually threading the coverslip collar up and down within the objective cap. A vacuum was connected to these ports using 1/16" ID PVC tubing connected to two straightflow rectangular manifolds with barbed tube adapters (McMaster-Carr 5233K51, 1023N13, and 4406T15, respectively). Pressure was increased slowly to stabilize capillaries without causing pain or tissue damage.

A flatfield correction algorithm was applied to the acquired images to correct for the intensity gradient across the field of view (Figure \ref{fig:Figure1}(c)-(d)). This was accomplished by dividing each image by a blurred version of itself \cite{Ford2012}. The blurred image was obtained by convolving with a Gaussian kernel with an 80-pixel standard deviation, which corresponds to 13 $\mu$m in object space. 

\FloatBarrier

\subsection{\textit{In Vivo} Phase Contrast Optimization}
Phase contrast in OBM relies on the net obliquity of scattered photons through the imaging plane, which increases with decreasing illumination-detection separation \cite{Ford2012}. In addition to this effect, we observe that, for very short separations, the backscatter of the illumination begins to overwhelm the phase contrast. To manage this tradeoff and optimize phase contrast, we measured phase contrast as a function of illumination-detection separation.

To optimize contrast to blood cells \textit{in vivo}, images of a single capillary were taken with the LED imaged critically on-axis, and the detector offset 208 $\mu$m (half of the FOV of the detector in object space) such that the peak intensity of the gradient field was conjugate to the edge of the sensor (Figure \ref{fig:Figure1}(c)-(d)). A video was acquired at 50 Hz of the same capillary as it was laterally translated from left to right across the field of view, while keeping the LED and sensor position fixed (Figure \ref{fig:Figure3}(a)-(d)). The phase contrast between adjacent, stacked red blood cells at different distances across the FOV was measured using Weber contrast ($C_{W} = (I_{max}-I_{min})/I_{min}$), calculated from plot profiles taken across the cells in the direction of the illumination-detection offset. Though phase contrast is most useful in imaging white blood cells, since these are rare, heterogenous, and difficult to visualize in the absence of phase contrast, we measured the phase contrast of the plasma membrane between adjacent, stacked red blood cells in order to characterize phase contrast over a range of illumination-detection separations.

\FloatBarrier

\subsection{\textit{In Vivo} Capillary Imaging}
The ventral tongue of human participants was imaged with approval by the Johns Hopkins University Institutional Review Board (IRB00204985). Three individuals were enrolled for imaging, with Fitzpatrick skin phototypes II, IV, and VI. Images of complete blood cell components (Figure \ref{fig:Figure4}) were selected from videos acquired at 200 Hz and 500 $\mu$s exposure time. Images of participants with different skin tone (Figure \ref{fig:Figure5}) were acquired at 50 Hz and 500 $\mu$s.

\subsection{Phase Contrast in Fixed Phantom}
In order to: (1) quantify phase contrast of the exact same phase object for a range of illumination-detection separations, and (2) verify the ability to reconstruct absorption-enhanced and phase-enhanced representations acquired from conventional oblique back-illumination microscopy, we imaged a phantom with fixed phase contrast. We constructed a back-scattering phase phantom by mixing 10$\%$ coffee creamer (French Vanilla Delight) in 90$\%$ water in a 1$\%$ w/v agar gel. The creamer creates lipid phase particles with low absorption within the sample. One such particle, approximately 10 $\mu$m in diameter, was imaged while the LED was laterally translated and the sensor was aligned on-axis. The Weber contrast was measured at each LED location by measuring the lateral intensity profile across the phase particle along the direction of LED motion. Additionally, images were taken at two diametrically opposed positions ($\delta x = \pm$240 $\mu$m), and either summed or subtracted, respectively, to create the characteristic absorption-enhanced and phase-enhanced images of oblique back-illumination microscopy \cite{Ford2012}.

\section{Results and Discussion}

\subsection{\textit{In Vivo} Phase Contrast Optimization}

\begin{figure}[h!] 
\centering\includegraphics[width=13.5cm]{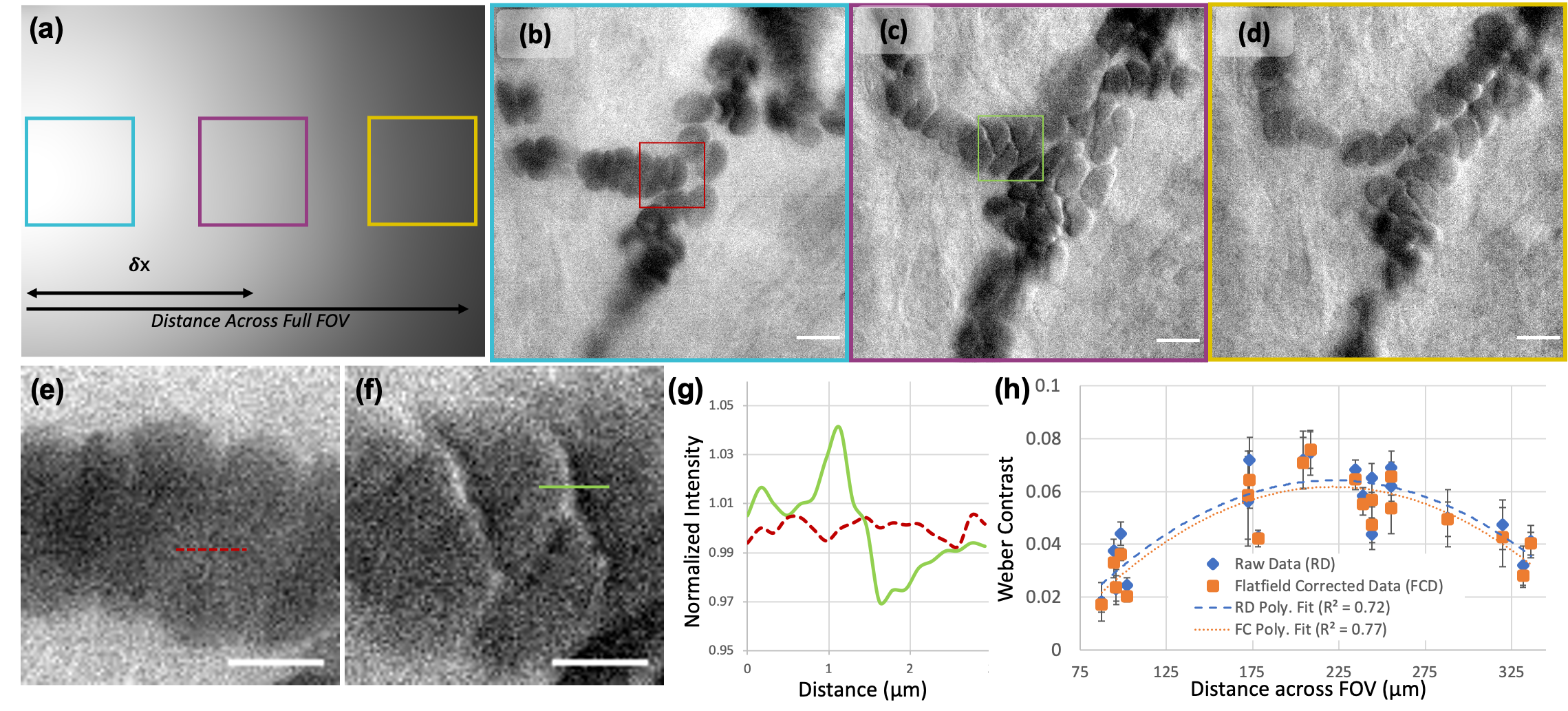}
\caption{Imaging a single capillary translated across the full FOV. (a) Full FOV with intensity gradient outlining where 90 x 90 $\mu$m regions of interest were acquired: cyan (b), magenta (c), and yellow (d). (b) Red blood cells close to the intensity peak (left edge of sensor) demonstrate only absorption contrast, (c) phase contrast is maximized near the center of the full FOV, and degrades due to low signal at large displacements, (d). (e) $\&$ (f) Enlarged regions of interest (15 x 15 $\mu$m) at locations of minimum and maximum phase contrast, respectively, with example normalized intensity plot profiles shown for each, (g). (h) Overall trend of distance across FOV vs. phase contrast demonstrates highest contrast at approximately 200-250 $\mu$m illumination-detection separation. Scale bars 10 $\mu$m (b)-(d) and 5 $\mu$m (e)-(f).}
\label{fig:Figure3}
\end{figure}

With the pneumatic objective cap turned off, a video was acquired of a single capillary in the ventral tongue of a human volunteer as it was swept laterally across the full FOV (Figure \ref{fig:Figure3}(a)-(d)). The plasma membrane of red blood cells appeared visible only from absorption contrast when the capillary was near the illumination axis (Figure \ref{fig:Figure3}(e)), and phase contrast was at a maximum near the center of the FOV (Figure \ref{fig:Figure3}(f)). Plot profiles across adjacent, stacked red blood cells show the improvement of phase contrast as the target is moved away from the peak illumination (Figure \ref{fig:Figure3}(g)). The Weber contrast between overlapping red blood cells was measured at known capillary distances across the FOV (Figure \ref{fig:Figure3}(h)). We found phase contrast was maximized at approximately 200-250 $\mu$m, regardless of whether the contrast was calculated with raw or flat field corrected data. Unless otherwise stated (as in the Appendix, section \ref{sec:Appendix}), all subsequent \textit{in vivo} data presented here was acquired with this geometry - the LED imaged on-axis, the sensor displaced 208 $\mu$m in object space, and capillaries stabilized near the center of the FOV where phase contrast was optimized.

\FloatBarrier

\subsection{\textit{In Vivo} Capillary Imaging}

\begin{figure}[h!] 
\centering\includegraphics[width=12cm]{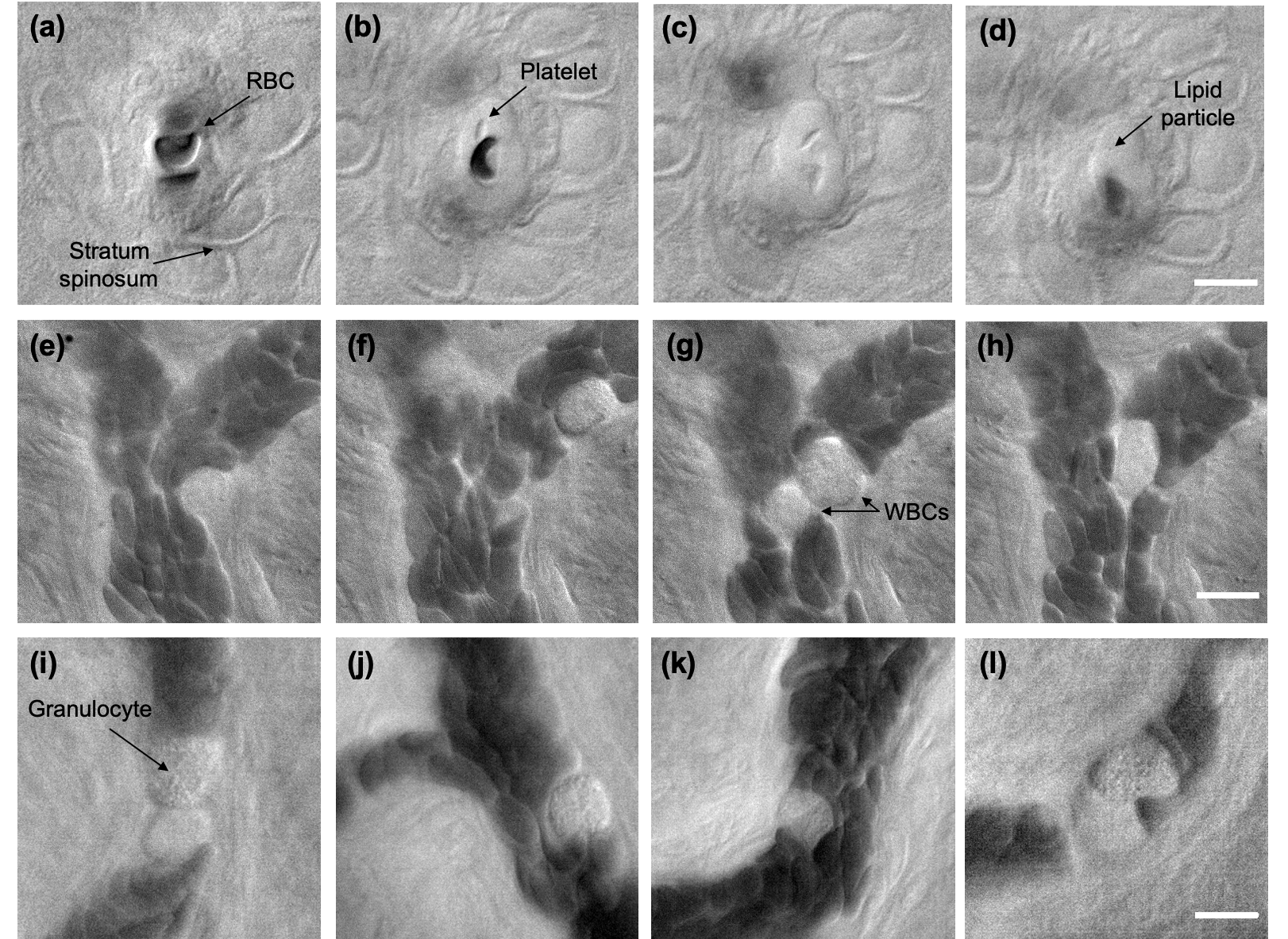}
\caption{Oblique back-illumination capillaroscopy can resolve red blood cells (RBCs), white blood cells (WBCs), platelets, and potentially chylomicrons or other lipid particles within human ventral tongue capillaries. (a)-(d) A superficial capillary within a dermal papilla surrounded by stratum spinosum shows red blood cells (a)-(b), platelets (b)-(c), and lipid particles (d) (see Visualization 1 and Visualization 2). (e)-(h) Thousands of red blood cells pass rapidly through larger vessels with intermittent white blood cells of various size and granularity (see Visualization 3). (i)-(l) Granular and agranular white blood cells of varying size are visible across multiple different capillaries (see Visualization 4 and Visualization 5). Scale bars are 10 $\mu$m.}
\label{fig:Figure4}
\end{figure}

Videos of capillaries and other larger blood vessels were acquired in the ventral tongue of a human participant with the LED imaged on-axis and the sensor offset. Capillaries were stabilized at the center of the field of view using suction from the pneumatic objective cap.  Figure \ref{fig:Figure4} shows example frames from these videos. Videos of these data can be seen in Visualizations 1-5.

Images in Figure \ref{fig:Figure4}(a)-(d) show a superficial capillary wide enough for only single red blood cell (RBC) passage. RBCs appear characteristically dark due to absorption of green light by hemoglobin. The outlines of RBC plasma membranes demonstrate the predicted phase contrast due to the offset sensor, critically imaged LED, and net oblique illumination. In plasma gaps between red blood cells, the phase contrast of the oblique back-illumination capillaroscope outlines platelets and other small particles approximately 600 nm in diameter that could be chylomicrons (Visualization 1 and Visualization 2). 

Images in Figure \ref{fig:Figure4}(e)-(h) highlight the presence of red and white blood cells passing through a larger vessel approximately 25 $\mu$m in diameter. For every several hundred RBCs that pass through a vessel, we observed a single or pair of WBCs with variable size and granularity. For example, the WBC in Figure \ref{fig:Figure4}(e) appears small and non-granular, 9.75 $\mu$m across, which are characteristics of a lymphocyte. The WBC in Figure \ref{fig:Figure4}(f) is 11 $\mu$m across and visibly granular, making it likely a granulocyte. Lastly, the WBC in Figure \ref{fig:Figure4}(h) is larger at 13.5 $\mu$m in length and non-granular, predicted features of a monocyte. Further work must be done \textit{in vitro} to isolate and image known examples of each WBC type before these labels can be confirmed.

\begin{figure}[h!] 
\centering\includegraphics[width=12cm]{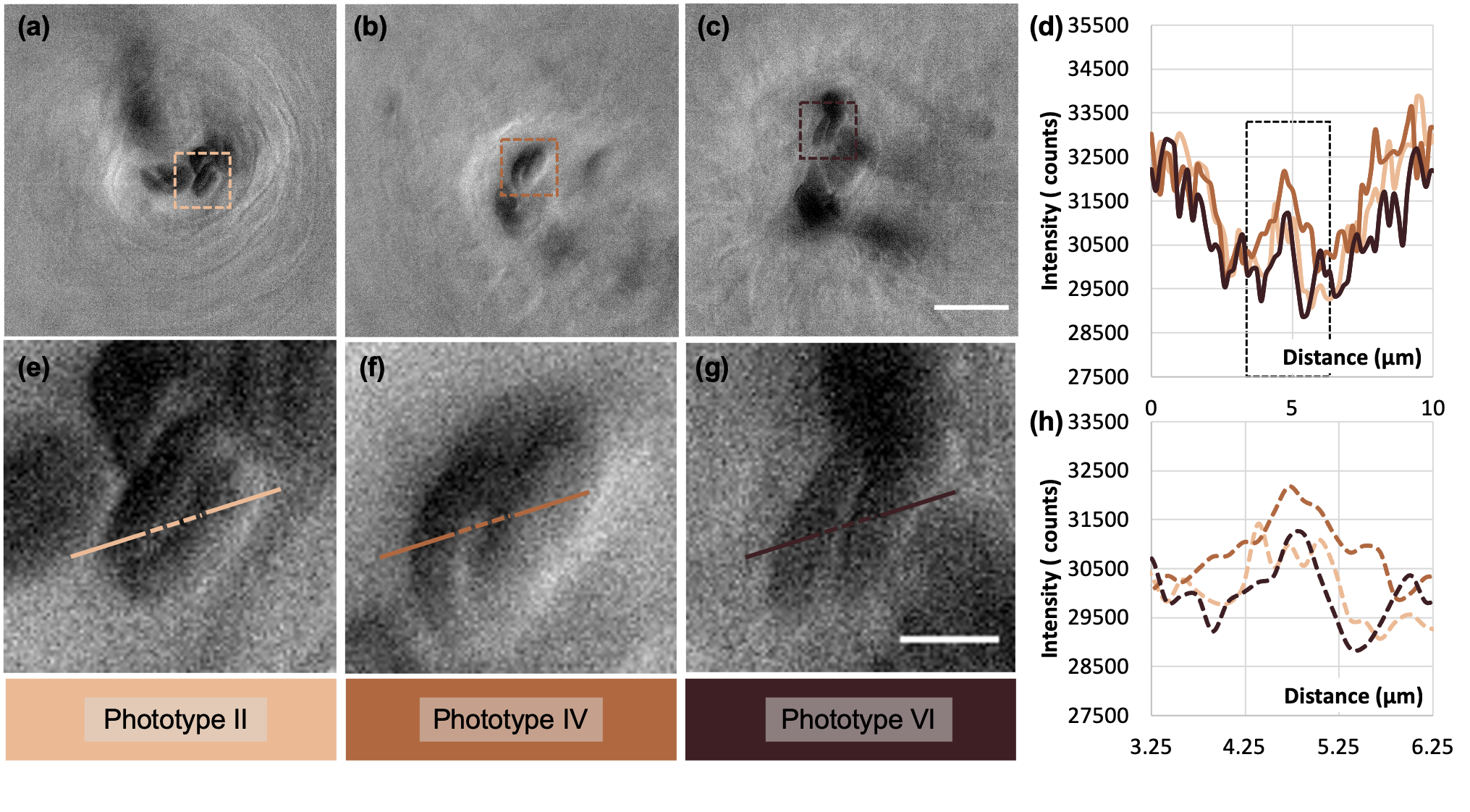}
\caption{(a)-(c) Imaging of three human subjects demonstrates robustness of ventral tongue imaging to differences in Fitzpatrick skin phototype. Scale bar 20 $\mu$m. (e)-(g) enlarged regions of interest with 5 $\mu$m scale bar. (d) Plot profiles across two adjacent stacked red blood cells in each image. (h) Enlarged plot profiles across adjacent red blood cell interface (dashed lines).}
\label{fig:Figure5}
\end{figure}

A device for non-invasive blood cell counting should work across individuals of different skin type. In this portion of the study, we enrolled three different individuals with Fitzpatrick skin phototype II, IV, and VI to test if contrast to the outline of red blood cells is similar despite differences in peripheral skin pigmentation. Figure \ref{fig:Figure5} shows the results of this experiment. As shown in Figure \ref{fig:Figure5}(h), the plot profile across two adjacent, stacked red blood cells demonstrates the predicted phase contrast pattern regardless of skin type. Weber contrast is 7.5$\%$, 8.1$\%$, and 7.9$\%$ for Fitzpatrick skin phototype II, IV, and VI, respectively.

\FloatBarrier
\FloatBarrier

\subsection{Phantom Phase Contrast Optimization}

\begin{figure}[h!] 
\centering\includegraphics[width=10cm]{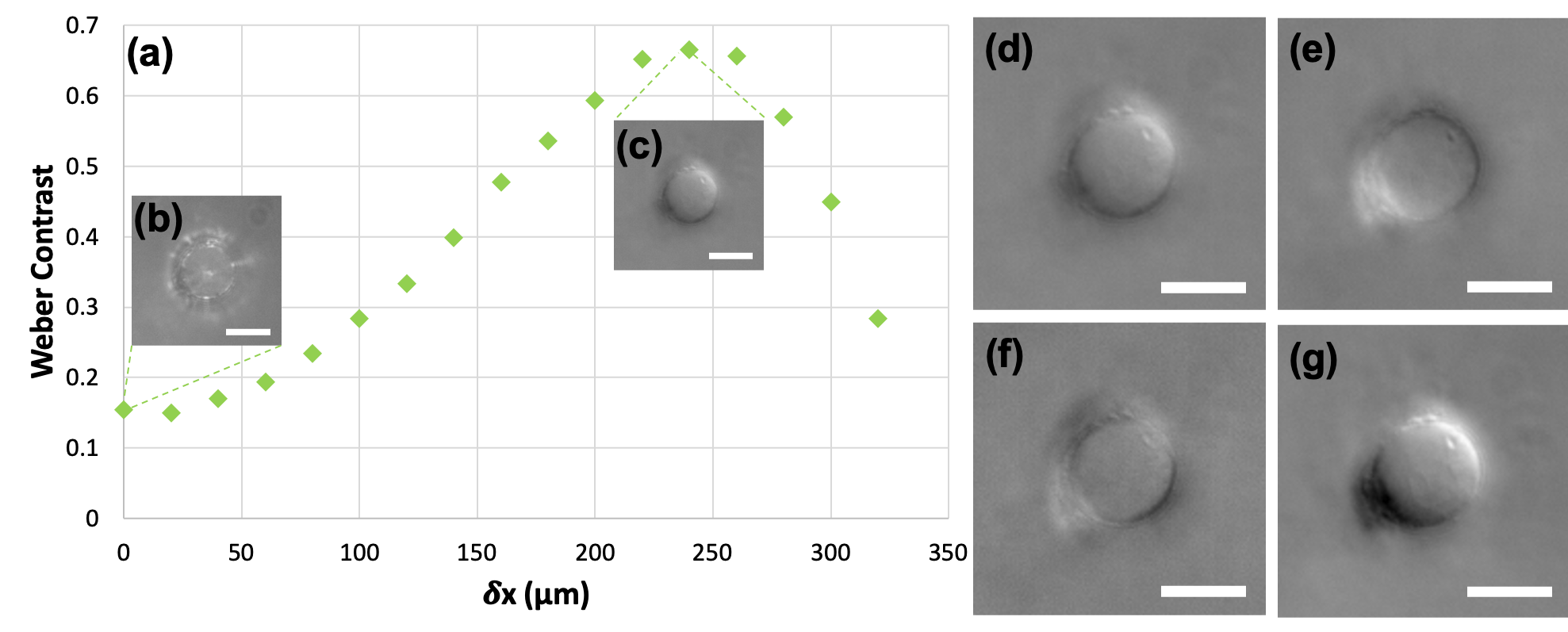}
\caption{(a) Images of a stationary 10 $\mu$m phase particle in a back-scattering phantom shows optimal phase contrast at 240 $\mu$m of LED displacement. Example images of this particle are shown for (b) on-axis LED imaging and (c) 240 $\mu$m LED displacement. (d), (e) When the illumination is diametrically opposed (±240 $\mu$m displacement), traditionally OBM allows the calculation of (f) absorption-only (addition image), and (g) phase-only (subtraction image) contrast. Scale bars 5 $\mu$m.}
\label{fig:Figure2}
\end{figure}

In an alternate geometry, illumination-detection separation can be created by keeping the sensor on axis, but laterally displacing the LED instead. To test the dependence of phase contrast on illumination-detection separation in this geometry, we utilized an agar phantom with creamer phase particles. We found that contrast was optimized with a conjugate LED displacement of approximately 240 $\mu$m in object space (Figure \ref{fig:Figure2}(a)), nearly consistent with the \textit{in vivo} phase contrast optimization results (Figure \ref{fig:Figure3}(h)). Images of the phase particle with on-axis and optimal off-axis illumination are shown as inserts in Figure \ref{fig:Figure2}(b)-(c). Images were also taken by translating the LED to two diametrically opposed positions ($\pm$240 $\mu$m) with respect to the phase particle (Figure \ref{fig:Figure2}(d)-(e)). Addition and subtraction of these two images generates absorption-only and phase-only images, respectively (Figure \ref{fig:Figure2}(f)-(g)). Together these images highlight the parallels between phase contrast produced in this technique and Fort et al.'s oblique back-illumination microscopy approach \cite{Ford2012}.

\FloatBarrier
\section{Conclusion}
Oblique back-illumination capillaroscopy in the ventral tongue is a simple, robust microscopy technique for imaging human blood cells regardless of skin tone. Combined with pneumatic stabilization, this technique is able to longitudinally image thousands of blood cells passing through individual human capillaries at high resolution and high speeds. We demonstrate that this technique allows visualization of red blood cells, white blood cells, and platelets, and sub-cellular granules are visible in white blood cells. This technique provides the basis for development of a non-invasive, label-free complete blood cell counter. A device with this capability would benefit a variety of patient populations including neonates, immunocompromised patients, and those in remote settings without access to conventional laboratory equipment and medical personnel required for a complete blood count. In its current form, ventral tongue oblique back-illumination capillaroscopy can be used as a research tool to easily and safely study a variety of hematologic and immunologic processes in humans, regardless of skin tone. 

\section{Appendix} 
\label{sec:Appendix}
\begin{figure}[h!] 
\centering\includegraphics[width=12cm]{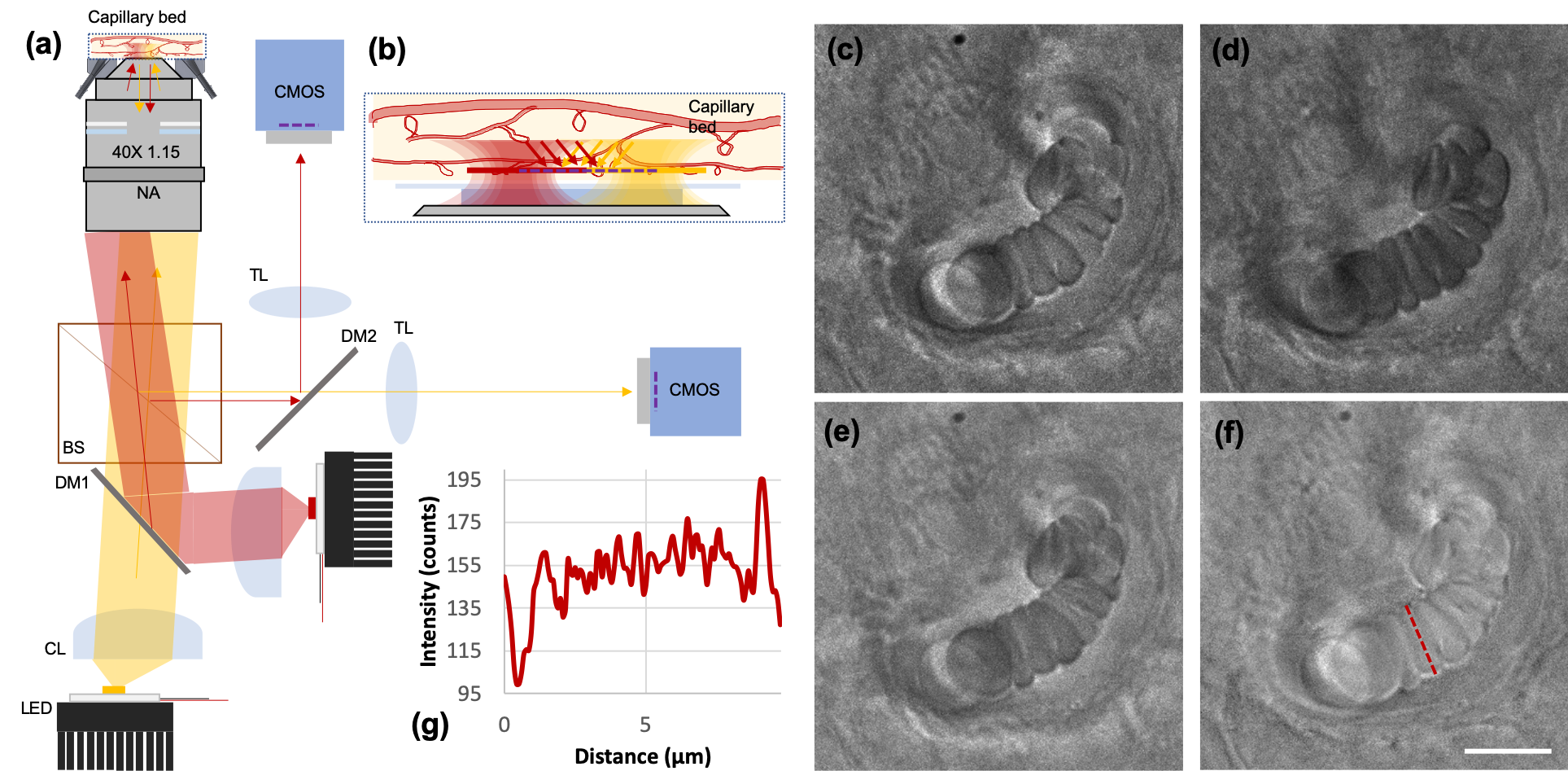}
\caption{(a) Schematic of dual source oblique back-illumination capillaroscope using diametrically opposed red and amber LEDs. The two sources are combined and later separated by two identical dichroic mirrors (DM) and projected onto two identical sensors. (b) Enlarged schematic of illumination technique in sample space where critically imaged, offset LEDs produce net diametrically opposed oblique illumination. (c) Red channel image, (d) amber channel image, (e) absorption-only (addition) image, (f) phase-only (subtraction) image, and (g) plot profile through red blood cell in phase-only image demonstrates characteristic phase contrast from oblique illumination. Scale bar is 10 $\mu$m).}
\label{fig:Figure6}
\end{figure}

Dual-source oblique back-illumination capillaroscopy is a natural extension of the proposed technique given prior publication on video-rate OBM \cite{Ford2013}. Figure \ref{fig:Figure6}(a)-(b) shows one possible realization of this technique with two, diametrically offset red and amber LEDs (Luxeonstar LXM5-PD01 and LXM5-PL01) combined and later separated using dichroic mirrors (DM) (Thorlabs DMLP605R). Two on-axis sensors (Imaging Source DMK33UX252) are used to simultaneously image the red and amber signal (Figure\ref{fig:Figure6}(c)-(d), respectively). Using the OBM technique of addition and subtraction, absorption-only and phase-only images can be generated (Figure \ref{fig:Figure6}(e)-(f), respectively). A plot profile across a red blood cell of the phase-only image demonstrates the characteristic contrast to the outline of the plasma membrane (Figure\ref{fig:Figure6}(g)). While an interesting extension, we did not find that the use of a dual-source oblique back-illumination capillaroscopy technique offered a sufficient increase in blood cell visibility to be worth the added complexity for this preliminary study. Further work could be done to optimize wavelengths for blood cell characterization and detection, and a quantitative phase reconstruction approach could be implemented as done with qOBM \cite{Obles2019}.

\FloatBarrier

\section*{Funding}
Johns Hopkins Medical Science Training Program Fellowship; NIH (R21 EB024700 and R43 EB024299).

\section*{Disclosures}
The authors are co-inventors on a provisional patent application assigned to Johns Hopkins University. They may be entitled to future royalties from intellectual property related to the technologies described in this article.

\bibliography{sample}

\end{document}